\documentclass[12pt,preprint]{aastex}
\begin{document}

\title{NICMOS SPECTROPHOTOMETRY AND MODELS FOR A--STARS}

\author{R.~C.\ Bohlin\altaffilmark{1} and Martin Cohen\altaffilmark{2}
\altaffiltext{1}{Space Telescope Science Institute, 3700 San Martin Drive,
Baltimore,  MD 21218; bohlin@stsci.edu}
\altaffiltext{2}{Radio Astronomy Lab, 601 Campbell Hall University of
California, Berkeley, CA 94720; mcohen@ astro.berkeley.edu}}
\slugcomment{To appear in AJ, 2008 September}
\received{2008 May 22}
\accepted{2008 July 3}
 
\begin{abstract}

Absolute flux distributions for eight stars are well measured from
0.8--2.5~$\mu$m with NICMOS grism spectrophotometry at a resolution of
R~$\sim100$ and an accuracy of 1--2\%. These SEDs are fit with Castelli \&
Kurucz model atmospheres; and the results are compared with the
Cohen-Walker-Witteborn (CWW) template models for the same stars. In some cases, the $T_\mathrm{eff}$, $\log g$, and $\log z$ parameters of the best fitting
model differ by up to 1000~K from the earlier CWW model. However, differences in the continua of the modeled IR flux distributions from 0.4--40~$\mu$m are always less than the quoted CWW uncertainty of 5\% because of compensating changes in the measured extinction. At wavelengths longward of the 2.5~$\mu$m NICMOS limit, uncertainties still approach 5\%, because A-star models are not yet perfect. All of these A~stars lie in the \emph{JWST} continuous viewing zone and will be important absolute flux standards for the 0.8--30~$\mu$m \emph{JWST} wavelength range.  
\end{abstract}


\section{Introduction}

The \emph{James Webb Space Telescope} (\emph{JWST}) is a NASA flagship mission and requires flux standards in its continuous viewing zones that are located near the ecliptic poles. The \emph{JWST} instrumentation covers the
0.8--30~$\mu$m region; and for reference flux distributions at the longer
wavelengths, stellar model atmospheres have traditionally been used to
extrapolate from the visible into the mid-IR. The nine A-star standards chosen
for \emph{JWST} are established by Cohen et~al.\ (2003a, CMHMS) as an extension of the original CWW standard star network (Cohen, Walker, \& Witteborn 1992; Cohen 2007). These nine stars are fainter than the \emph{JWST}/NIRSpec saturation limit and have already been observed by the Infrared Array Camera (IRAC) on the \emph{Spitzer Space Telescope}; two of these nine (HD~165459 and 1812095) are primary standards used for the IRAC flux calibration (Reach et~al.\ 2005).

In order to compare and cross calibrate \emph{HST} and the IRAC \emph{Spitzer} absolute flux scales, the nine \emph{JWST} standards have been observed by the NICMOS objective grism spectrometer in the 2006--2008 time frame. NICMOS spectra at a resolution R$\,\sim$100 are obtained in three modes G096, G141, and G206, which together cover the 0.8--2.5~$\mu$m range. The absolute flux calibration of these spectra is established on the \emph{Hubble Space Telescope} (\emph{HST}) white dwarf (WD) flux scale by observations of the primary pure hydrogen WD stars GD71, GD153, and G191B2B (Bohlin 2000, Bohlin, Dickenson, \& Calzetti 2001, Bohlin 2003, Bohlin \& Koester 2008) in these three NICMOS modes. The NICMOS count rate
spectrophotometry is extracted from the 256$\times$256 images per the prescription of Bohlin, Lindler, \& Riess (2005). The ratios of the count rate spectra for the A--stars to the count rates for the primary WD flux standards determine the absolute fluxes of the A--stars after correcting for the non-linearity of the NICMOS HgCdTe array (Bohlin, Riess, \& de~Jong 2006; Bohlin 2007). The fluxes for the three NICMOS modes are combined into one measured SED covering the full 0.8--2.5~$\mu$m range. The modeled flux distributions of the  primary WDs are defined by the stellar temperatures and gravities, which are derived from the Balmer line profiles. As a typical example, the uncertainty of 3000~K in the 61193~K effective temperature of G191B2B means that its relative flux should be correct to better than 1\% from 0.35 to 2.5~$\mu$m.

\section{Comparison between NICMOS and CWW Fluxes}

Figure 1 shows the ratio of the NICMOS to the CWW fluxes binned to a resolution
of R~$=100$. The red circles are the ratio of the \emph{BVRIJHK$_{s}$}
photometry to the CWW fluxes as integrated over the photometric bandpasses. The bandpass functions are  corrected for atmospheric transmission and are from
CMHMS for the Mount Hopkins Observatory (MHO) \emph{BVRI} photometry and from Cohen  et~al.\ (2003b, CWM) for the 2MASS \emph{JHK$_{s}$} photometry (Skrutskie et~al.\ 2006). The photometry on the Landolt-Johnson-Cousins scale is from the headers of the individual CWW flux templates and is summarized in Table~1. There is also \emph{BVRI} photometry for three stars from independent observations at the Instituto de Astrofisica de Canarias (IAC); but the IAC data are not included here for consistency. The average MHO and IAC values differ by a maximum of 0.047~mag, consistent with the adopted 1$\sigma$ uncertainty for the MHO photometry of 0.03~mag. CMHMS also utilized the available \emph{Tycho} and \emph{Hipparcos} photometry. HD~165459 is too bright for the MHO/IAC programs, so that the $B$ and $V$ photometry is from Simbad.  

The photometry zero points are defined by the composite STIS flux plus a 9400~K
Kurucz model named $alpha\_lyr\_stis\_003.fits$ (Bohlin \& Gilliland 2004; Bohlin 2007) from the CALSPEC database\footnote{http://www.stsci.edu/hst/observatory/cdbs/calspec.html/.} and by the
adopted photometry for Vega from Ma\'{\i}z Apell\'aniz (2007), i.e., 0.034,
0.026, 0.030, 0.017, $-0.021$, 0.009, and 0.000 at $B$, $V$, $R$, $I$, $J$, $H$,
and $K_{s}$, respectively. These photometric zero points agree with those in
CMHMS and CWM to 0.02~mag and to 0.01~mag for $B$, $V$, and $R$. Holberg \& Bergeron (2006) also agree to $<$0.02~mag, except for their $J=+0.023$ value.

All of the NICMOS flux measurements and ground based photometry agree with the CWW SEDs within the quoted 5\% accuracy of the CMHMS A-star template flux distributions. However, with a goal of 1--2\% for the final standard star
fluxes, improvements in fitting models to the data must be pursued. Immediately
obvious in Figure~1 is a feature near 1.5~$\mu$m for all nine stars that has a
1.55~$\mu$m peak to 1.47~$\mu$m valley difference of $\sim$7\%. This small
discrepancy is caused by deficiencies in the early Kurucz (1993a, 1993b) models
used for the CMHMS templates at the convergence of the hydrogen Brackett lines
to the Brackett continuum (Bohlin 2007). Figure~2 compares the same observations to the improved models of Castelli \& Kurucz (2004 CK04), where the Brackett opacity is properly computed. These models have a microturbulent velocity of 2~km/s and do not include convective overshoot. In Figure~2, the NICMOS flux for each star is divided by the CK04 model with the same $T_\mathrm{eff}$, $\log g$, $\log z$, and color excess $E(B-V)$ used for the CWW template in Figure~1. Each reddened model is normalized to the observed NICMOS flux, so that the average ratio has the value of 1.0000 that is written in each panel of Figure~2. Our reddening curve is from Cardelli, Clayton, \& Mathis (1989, CCM) at wavelengths shorter than 1.2~$\mu$m and at longer wavelengths from Chiar \& Tielens (2006) after matching to CCM at 1.2~$\mu$m. Thus, the discrepant 1.5~$\mu$m features disappear, and there are no average differences between the observed and modeled fluxes. The residual rms differences between the NICMOS flux and the normalized model are reduced for all nine stars. Because the H~\textsc{i} lines are poorly sampled in the CK04 grid where a single point often defines the total equivalent width of the line, the only valid ratio points lie in the continuum regions defined in Table~2. The average ratio and rms values appearing for each star in Figures~1--2\pagebreak\ represent the average and scatter of the NICMOS data points in these continuum regions only, while the large open circles represent the average
ratio in each of these eight continuum regions.

In addition to the above $\sim$7\% improvement to the original CWW Kurucz models at the Brackett convergence, the new CK04 models have weaker absorption by $\sim$3\% at the convergence of the Paschen lines. No corresponding change appears at the Balmer convergence, where the original physics must have been adequate. There is about a 2\% difference near the Pfund convergence at 2.3~$\mu$m. However from 0.3--40~$\mu$m, the only other differences that exceed 2\% between the Kurucz models used for the CWW templates and the new CK04 models are narrow band differences in a few H~\textsc{i} lines. The new CK04 models differ little in the broad photometry bands, as verified by the relative positions of the red data points between the comparison with the CWW models in Figures~1 to the comparison with the CK04 models in Figure~2. To a precision of $\sim$1\%, the amount of the shift in the photometry points is the same as the shift in the average NICMOS differences from the values written in Figure~1 to unity in Figure~2.

\section{What Are the Expected Residuals?}

The residual rms differences between the NICMOS fluxes and the CK04 models range from 0.84--1.81\% in Figure~2. If the models were a perfect representation of the SEDs, the rms residuals should be comparable to those of a star with about the same flux as observed in a similar time allocation of one or two orbits. The primary standard star G191B2B is as bright as the fainter A stars of Figure~2. In comparison to the pure hydrogen NLTE model for G191B2B, the residuals for the four NICMOS spectra of G191B2B obtained in four separate two-orbit observation programs range from 0.5--0.8\% over the 0.82--2.4~$\mu$m region. Thus, the best explanation for larger residuals in Figure~2 is that the chosen models are not good enough representations of the true IR stellar fluxes. The NICMOS observations are all obtained in one orbit; but the expected rms is not as much as 1.4$\times$ worse, because errors in the flat field dominate the residuals for such bright stars.

The possibilities for poor models include either a deficient modeling code or
the wrong physical parameters for the model chosen to represent the actual star.
How realistic are the models of the CK04 grid in the A~star range? For Vega
(A0V) in the visible, Bohlin (2007) demonstrated excellent agreement between the
R$\,\sim500$ STIS spectrum and a special high fidelity Vega model from the Kurucz website, e.g., rms differences of $<$0.3\% from 4500--8200~\AA. Figure~3 compares two Kurucz high fidelity models to the corresponding models
interpolated from the CK04 grid. Agreement of the CK04 models at 9400~K and
9550~K with the hi-fi models is nearly perfect in the continuum with deviation
rarely exceeding 1\% in the lines, where the sparse wavelength sampling and
averaging over the sampling intervals causes small errors in the CK04 lines.
Because the main differences between the CK04 grid and the high fidelity models
are a finer sampling in wavelength and better preservation of proper equivalent widths, this good agreement is expected; the physics is the same and interpolation within the CK04 grid is accurate. For the zero points established
by the 9400~K hi-fi model, differences with the corresponding lo-fi CK04 model
are $<$$\sim$0.1\% in the broad photometry bands (red points). Thus, the
remaining likely explanation of the larger than expected residuals in Figure~2
is that the CWW model parameters are not the best choices to fit the NICMOS
observed flux distributions.

\section{Model Parameters for the Best Fit to NICMOS Fluxes}

\subsection{Detailed Example for 1812524}

In Figure~2, one of the poorest fits with the CWW stellar parameters is 1812524.
For this star, Figure~4 compares the 1812524 observations with various CK04
models from the CWW parameters in the bottom panel~(a) to the best fit in the
top panel~(e). The reddening is adjusted to minimize the average NICMOS
residuals in the continuum bands of Table~2. Fitting the NICMOS fluxes tends to
increase the residuals in the $B$, $V$, and $R$~bands. In the bottom panel,
increasing $E(B-V)$ slightly from the CWW value of 0.14 used in Figure~2 to 0.18
reduces the NICMOS rms residuals from 1.70\% to an acceptable 0.8\%. However, the fit to the \emph{BVR} photometry shows a 13\% error at~$B$, i.e., a 4$\sigma$ difference per our adopted 1$\sigma$ uncertainty in the MHO photometry of 3\%.

The best fit to the NICMOS flux distribution shown in panel~(b) reduces $\sigma$
to 0.60\%, while also improving the fit in the \emph{BVR} bands. This best fit is for $T_\mathrm{eff}=7800$~K and $\log g=3.0$, which corresponds to a spectral type near A8II (de~Jager \& Nieuwenhuijzen 1987). 

Spectral classification observations from the CWW program at Mount Hopkins
Observatory (MHO) exist for all nine of our stars. For 1812524, the metal line
strengths are near the limiting signal-to-noise; but the Balmer line profiles rule out the classification of A8II that is implied by the best fit to the NICMOS flux distribution. The observed Balmer lines are too strong for luminosity classes II--III near A8, but suggest a main sequence classification in the A1--A7 range, i.e., $T_\mathrm{eff}$ in the range 7800--9400~K. For this spectral range, de~Jager \& Nieuwenhuijzen have $\log g$ in the 4.2--4.3 range for a main sequence classification. Assigning an uncertainty of half way between luminosity III and V implies a lower limit of 4.0 for $\log g$. In the middle panel~(c) of Figure~2, $\log g$ has this minimum allowable 4.0 value and a $\sigma=0.61\%$. Even though the NICMOS residuals are not significantly increased over the best fit in panel~(b), the \emph{BVR} photometry residuals become unacceptably large.

In order to find an acceptable fit to both the \emph{BVR} and NICMOS data, the
\emph{BVR} residuals are included in the total residual to be minimized in the
top two panels. However, the rms residuals written near the bottom of each panel
remain as NICMOS values only, in order to compare among all five panels.
Panel~(d) produces reasonable \emph{BVR} residuals at the expense of a poorer
fit to the NICMOS spectrophotometry. Finally in the top panel, reducing the
heavy element abundance, $\log z$, to the optimum $-1.0$ produces a $\sigma=1.0\%$. Unfortunately, the constraints on this best-fitting model produce a poorer rms fit than expected from the typical G191B2B residual scatter. Furthermore, the NICMOS residuals show a trend of increasing with wavelength, while the optical \emph{BVR} residuals are also high. This pattern
would be expected, if 1812524 is an unresolved binary consisting of one early
A~star and one cooler star. Shorter wavelength fluxes along with mid-IR
\emph{Spitzer} IRAC and MIPS observations are needed to address the question of multiplicity.

Figure~5 illustrates the ratio of the CWW absolute fluxes to the new result for
the cooler temperature of $T_\mathrm{eff}=8450$~K and the compensating lower reddening of $E(B-V)=0.067$. Flux differences bigger than 5\% occur only below the Balmer jump. UV observations with the revived STIS following the final \emph{HST} servicing mission are required to fully verify and refine the newly derived $T_\mathrm{eff}$ and extinction for 1812524. The 0.22~$\mu$m dust extinction feature will provide a strong constraint on $E(B-V)$. The new model agrees with the CWW template typically within 1\% from two to 40~$\mu$m. However if 1812524 is a double star, neither model is a reliable predictor of the mid-IR fluxes; and neither model should be used to predict absolute fluxes.

\subsection{Best Fits for the Other Eight Stars}

Following the technique used above for 1812524, values for $T_\mathrm{eff}$,
$\log g$, $\log z$, and $E(B-V)$ are derived for the remaining eight stars in
the NICMOS-\emph{JWST} standards program. Just as for 1812524, the classification spectra limit $\log g$ to a minimum of 4.0, except for 1743045, where the Balmer lines are much weaker than for the other eight stars. Figure~6 illustrates the residuals for the NICMOS fluxes divided by the best fitting model for all nine program stars; and these results are summarized in Table~1. The best fits are derived by minimizing the residuals when the \emph{BVR} photometry is included along with the NICMOS data. The rms residual $\sigma$ values for just the NICMOS fluxes are written on the plots for comparison with the spread of residuals for the pure hydrogen WD~G191B2B. 

In the continuum regions of Table~2, nearly all of the individual NICMOS data
points lie within 2\% of the best fitting model; and the residual rms values are
all $<$0.82\%, except for 1739431 and 1812524. However, the 1739431 spectral
images have a FWHM about 20\% larger than normal in the direction perpendicular to the dispersion. Either these data are out of focus or 1739431 is a multiple system. Because of the wavelength dependence of the steep peak-to-valley structure in the NICMOS sensitivity functions, especially for G096 (Bohlin, Riess, \& de~Jong 2006), the derived fluxes are very sensitive to focus or to an anomalous mixture of wavelengths on a pixel. Thus, the results for 1739431 are suspect, and neither the observed NICMOS fluxes nor the fitted model should be used as a flux standard. 

\subsection{Consistency Check}

In comparison to the CWW and CMHMS stellar parameters, the generally cooler
$T_\mathrm{eff}$ found in this work imply later spectral types, and redder
intrinsic color $(B-V)_{o}$, as shown in Table~1. The MK types are estimated
from the tabulations in de~Jager \& Nieuwenhuijzen (1987) by picking the nearest
MK type corresponding to the newly derived $T_\mathrm{eff}$ and $\log g$ in
Table 1. Intrinsic colors are from Landolt-B\"ornstein (1982). Subtracting the
intrinsic $(B-V)_{o}$ from the measured MHO $(B-V)$ provides another estimate of the color excess $E(star)$ in the final column of Table~1. From a comparison to other spectral type calibrations (e.g., Gray 1992), there is an uncertainty of about two spectral sub-classes in the spectral type vs.\ $T_\mathrm{eff}$ near type A6, which corresponds to an uncertainty in $(B-V)_{o}$ of $\sim$0.05~mag. The color excess $E(fit)$ from the best fitting model agrees with the traditional estimate $E(star)$ to 0.02~mag for all eight A-star standards with valid observations. Even though the CWW color excess $E(B-V)$ differs from the new best $E(fit)$ by 0.07--0.09~mag for five of the eight stars, the newly derived model parameters, reddening, and spectral types form an internally consistent picture where both independent estimates of the color excess agree within 0.02~mag for all eight stars. The new NICMOS spectral types all agree with the Cohen types within three sub-classes, which is consistent with a re-examination of the CCW classification spectra.

CMHMS first derived estimates of their spectral types classically, by measuring
optical spectral features and comparing with those of MK standards; and then the
most consistent combination of this type, the reddening, and the photometry was
derived for each star. By contrast, the method described in this paper is reversed, proceeding from the best fitting synthetic spectrum from an updated
set of models to the spectral type implied by the $T_\mathrm{eff}$ and $\log g$
of the best fit. Despite these differences of philosophy and technique,\pagebreak\ the resulting energy distributions that emerge are robust. See the Error Analysis section below for quantification of the differences between the two sets of SEDs.

\subsection{The 2MASS and MHO Photometry}

Most of the 2MASS \emph{JHK$_{s}$} photometry in Figure~6 agree with the NICMOS data within the quoted 2MASS uncertainties of 2--3\%. An exception is in the $K_{s}$ band for 1743045, where the 2MASS point is high by 2.5$\sigma$, per the 2MASS $\sigma=0.028$~mag. Also, the $J$~band for the 1739431 is low by 2--3$\sigma$. 

With a one $\sigma$ uncertainty of 3\% in the optical, the $B$, $V$, $R$, and
$I$ MHO photometry is in excellent agreement with the new models; only for the
poorly fitting model for 1812524 do any of the measurements differ by as much as
4\% from the synthetic photometry from the models. 

\subsection{Error Analysis}

Even though our revised model parameters often differ substantially from those
of CWW, the model flux distributions are still similar, because lower reddening
partially compensates for lower temperatures. Other than the few percent
improvements in the modeling of the Paschen, Brackett, and Pfund
line-convergence regions, the largest difference in the IR from 0.8--40~$\mu$m
between the CWW and the new SEDs is for 1743045, where the new model is up to 4\% fainter from  0.9--1~$\mu$m and $\sim$3\% fainter from 3.3--8~$\mu$m. With coverage from 3--9~$\mu$m, the IRAC data for 1743045 may better fit one model or the other. The published IRAC data for HD~165459 and 1812095 (Reach  et~al.\ 2005) do not strongly favor one model or the other, because the models agree to $<$1\% for HD~165459 and to $<$2\% for 1812095 over the 3--9~$\mu$m IRAC range. When the IRAC data for all eight stars are available, a separate paper is planned, which will utilize the revised calibration of Rieke, et~al.\ (2008). Data at longer wavelengths provide a more sensitive check for excess emission from dust rings like the one around Vega; but \emph{JWST} observations can identify discrepantly large cases of excess emission longward of 10~$\mu$m, if no other mid-IR data are available before \emph{JWST} operations begin.

For the seven stars with good models, the generally cooler $T_\mathrm{eff}$
found in this work implies a systematic larger flux than the CWW/CMHMS SEDs of
$\sim$0.5\% at 12~$\mu$m, which increases monotonically to $\sim$1.5\% at
40~$\mu$m. However, the typical $\sim$2\% agreement between the old CWW models and the new fits to the NICMOS data is not a good measure of the uncertainty in the extrapolation of the absolute fluxes longward of the 2.5~$\mu$m NICMOS limit. There could be systematic uncertainties in the Kurucz heritage code used to calculate both the CWW and CK04 models. 

Only if entirely different computer codes produce the same flux distributions
for the same $T_\mathrm{eff}$, $\log g$, and $\log z$ would confidence in the
mid-IR fluxes approach 2\%. A small search for independent models was conducted, which included those with heritage from B.~Plez (MARCS), P.~Hauschildt (NextGen Phoenix), and T.~Lanz. The MARCS models are only for $T_\mathrm{eff}<8000$~K (Plez 2008, priv.\ comm.). A grid of Phoenix models is
available\footnote{ftp://ftp.hs.uni-hamburg.de/pub/outgoing/phoenix/GAIA/.}
(Brott \& Hauschildt 2008, priv.\ comm.); and the model appropriate for Vega
with $T_\mathrm{eff}=9400~K$, $\log g=4$, and $\log z=-0.5$ is compared with the STIS observations of Vega (Bohlin \& Gilliland 2004, Bohlin 2007) and with the Kurucz Vega model for $T_\mathrm{eff}=9400~K$, $\log g=3.9$, and $\log z=-0.5$. While there is excellent agreement between the STIS spectrum and the Kurucz model, many absorption lines are far too strong in the Phoenix model; e.g., lines at 4250 and 4406~\AA\ have equivalent widths of 1--2~\AA\ in the Phoenix model but are $<$0.1~\AA\ in the observation and Kurucz model.

The most helpful comparison models are provided by T.~Lanz (2008, priv.\ comm.), who started with an Atlas~9 model but used the Hubeny \& Lanz Synspec code\footnote{http://nova.astro.umd.edu/.} to compute the detailed spectral distributions. One model appropriate to Vega ($T_\mathrm{eff}=9400$~K, $\log g=3.9$, and $\log z=-0.5$) agrees well with the Kurucz 9400~K model, i.e., within 2\% from 1--40~$\mu$m. In the narrow region of the Balmer convergence from 3670--3870~\AA\ where the physics is not yet perfected, the Lanz Synspec model improves the fit to the STIS observations of Vega by up to 5\% between the Balmer lines but is not as good as the special 9400~K Kurucz model continuum near 3700~\AA. For a cooler Synspec model ($T_\mathrm{eff}=8020~K$, $\log g=3.7$, and $\log z=-1.5$), the agreement with the interpolated CK04 model is not as good in the IR. Differences of up to 4\% at 10~$\mu$m are present, perhaps due to different treatments of free-free opacity. Thus, a conservative approach is to adopt the same 5\% uncertainty as CMHMS for the newly modeled IR fluxes at wavelengths longward of 2.5~$\mu$m and from 0.4--0.8~$\mu$m. In the observed NICMOS range of 0.8--2.5~$\mu$m uncertainties are estimated to be 2--3\%. Shortward of the Balmer limit, the model fits are not constrained by observation and should be used only for rough flux estimates. STIS observations from a short wavelength limit of 1150~\AA\ are required to establish good UV flux distributions and refine the CK04 model fits in the \emph{BVR} photometry range.

\section{Summary}

Absolute flux standards are required for \emph{JWST} calibration in the
\emph{JWST} continuous viewing zone near the north ecliptic pole. NICMOS grism observations measure the flux distributions of eight A~stars, seven of which fit CK04 model atmospheres from 0.82--2.4~$\mu$m within an rms scatter of $<$0.8\%. These models are normalized to the NICMOS SEDs and are used to extend the measured NICMOS fluxes to shorter and longer wavelengths. These seven composite NICMOS plus model SEDs are archived in the CALSPEC database of \emph{HST} flux standards\footnote{http://www.stsci.edu/hst/observatory/cdbs/calspec.html/} as \emph{*\_nic\_002.fits}. The eighth well measured NICMOS flux distribution for 1812524 is also included in the same CALSPEC web page as \emph{1812524\_nic\_002.fits} but does not have its wavelength range extended beyond the NICMOS range. 

\acknowledgments

G.~Kriss and J.~Rhoads motivated the NICMOS observations and assisted in the 
proposal process along with R.~Diaz-Miller. J.~Valenti and R.~Kurucz provided
helpful comments on an early draft of this paper. D.~Lindler analyzed the poorly
focussed NICMOS spectra of 1739431 and supported the IDL spectral extraction
package. Support for this work was provided by NASA through the Space Telescope Science Institute, which is operated by AURA, Inc., under NASA contract NAS5-26555. This research made use of the SIMBAD database, operated at CDS, Strasbourg, France. This publication makes use of data products from the Two Micron All Sky Survey, which is a joint project of the University of Massachusetts and the Infrared Processing and Analysis Center/California Institute of Technology, funded by the National Aeronautics and Space Administration and the National Science Foundation.

\newpage
\begin{deluxetable}{lccccccccccccccccc}
\rotate
\tabletypesize{\scriptsize}
\tablewidth{0pt}
\tablecolumns{19}
\tablecaption{Stars Observed with the NICMOS Grisms}
\tablehead{
\colhead{Star} &\colhead{$B$} &\colhead{$V$} &\colhead{$R$} &\colhead{$I$}
&\colhead{$T_\mathrm{eff}$} &\colhead{$\log g$} &\colhead{Sp.T.} &\colhead{$E(B\!-\!V)$}
&&\colhead{$T_\mathrm{eff}$} &\colhead{$\log g$} &\colhead{$\log z$} &\colhead{$E(fit)$} 
&\colhead{Sp.T.} &\colhead{$(B\!-\!V)_o$} &\colhead{$(B\!-\!V)$} &\colhead{$E(star)$}\\
&\multicolumn{8}{c}{CWW} &&\multicolumn{8}{c}{This work}
}
\startdata
HD165459 &\phn6.994\rlap{\tablenotemark{a}} &\phn6.864\rlap{\tablenotemark{a}} &\nodata &\nodata &9397 &4.18 &A1V &0.09 &&8600 &4.20 &$-1.5$ &0.017 &A4V &0.12 &0.13 &0.01\\
1732526 &12.647 &12.530 &12.474 &12.407 &8710 &4.21 &A3V &0.04 &&8500 &4.00 &$-0.5$ &0.023 &A4V &0.12 &0.12 &0.00\\
1739431\tablenotemark{b} &12.505 &12.311 &12.225 &12.129 &8710 &4.21 &A3V &0.10 &&8500 &4.00 &$-1.5$ &0.079 &A4V &0.12 &0.19 &0.07\\
1740346 &12.678 &12.478 &12.381 &12.271 &8185 &4.25 &A5V &0.06 &&8050 &4.00 &$-1.5$ &0.032 &A6V &0.18 &0.20 &0.02\\
1743045 &13.803 &13.525 &13.378 &13.223 &8185 &4.25 &A5V &0.14 &&7650 &3.80 &$-1.0$ &0.049 &A8III &0.25 &0.28 &0.03\\
1802271 &12.065 &11.985 &11.978 &11.955 &8710 &4.21 &A3V &0.00 &&9100 &4.00 &$-0.5$ &0.024 &A2V &0.05 &0.08 &0.03\\
1805292 &12.413 &12.278 &12.230 &12.164 &9397 &4.18 &A1V &0.10 &&8400 &4.00 &$-1.0$ &0.006 &A4V &0.12 &0.14 &0.02\\
1812095 &11.941 &11.736 &11.632 &11.526 &9016 &4.20 &A2V &0.13 &&8250 &4.05 &$-1.5$ &0.043 &A5V &0.15 &0.20 &0.05\\
1812524 &12.455 &12.273 &12.191 &12.103 &9397 &4.18 &A1V &0.14 &&8450 &4.00 &$-1.0$ &0.067 &A4V &0.12 &0.18 &0.06\\
\enddata
\tablenotetext{a}{Simbad. Simbad has A2 for HD~165459.}
\tablenotetext{b}{Results for 1739431 are unreliable, because the NICMOS observations are out of focus.}
\end{deluxetable}

\begin{deluxetable}{c}
\tablewidth{0pt}
\tablecaption{IR Continuum Regions}
\tablehead{
\colhead{Wavelength Range ($\mu$m)}
}
\startdata
\phn0.81--0.845\\
\phn0.97--0.995\\
1.02--1.08\\
1.11--1.26\\
1.30--1.55\\
\phn1.75--1.805\\
1.97--2.15\\
\enddata
\end{deluxetable}

\clearpage

\begin{figure}  
\epsscale{.8}
\plotone{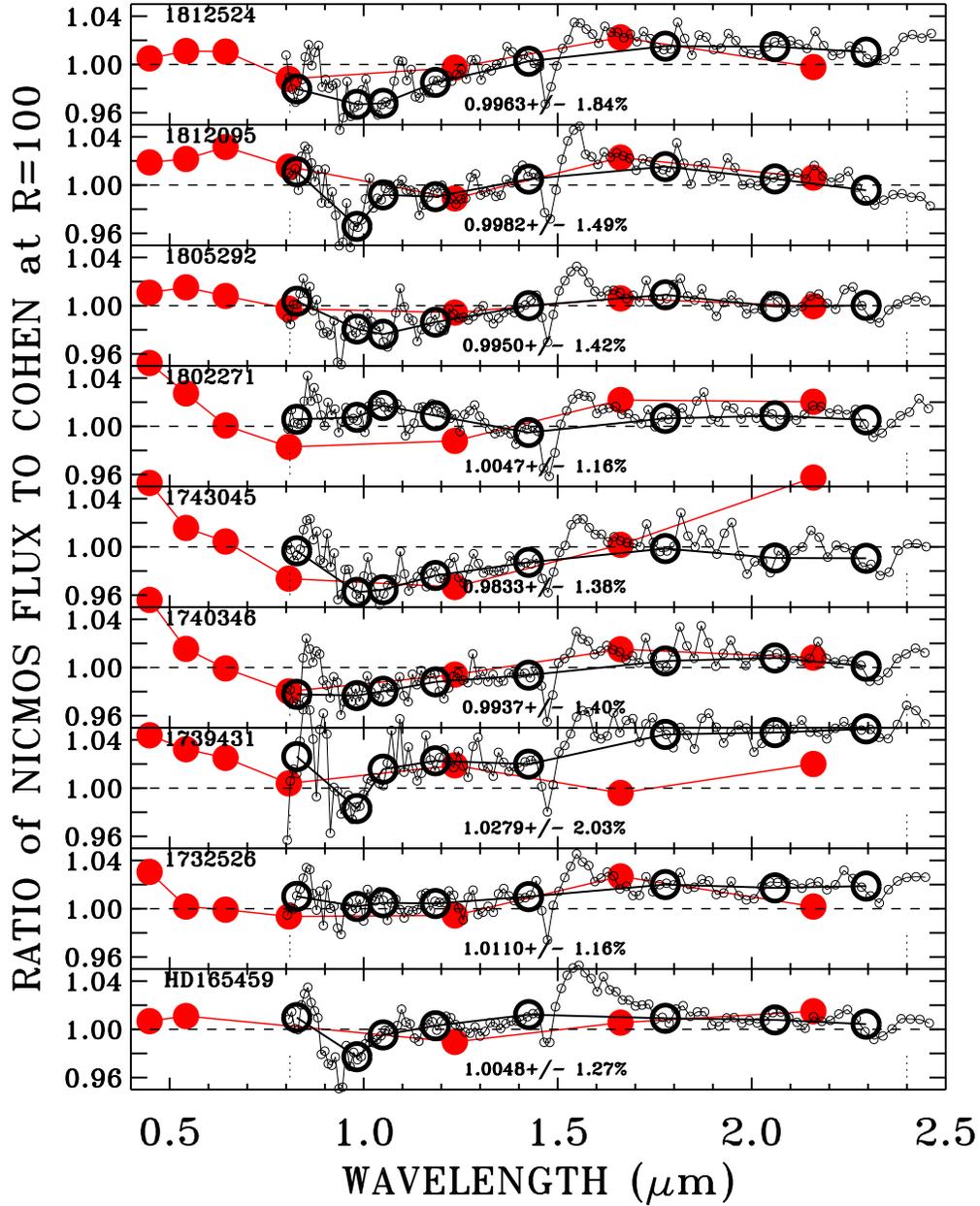}
\caption{
The ratio of the NICMOS absolute fluxes to those provided by M.~Cohen from the
CWW network of standard stars (small open black circles) at a resolution of
R~=~100. The large open black circles are the average ratios in each of the
eight bandpass regions of Table~2. The filled red circles are similar ratios for
\emph{BVRIJHK$_{s}$} photometry, where the denominator CCW fluxes are integrated over the bandpass functions provided by CMHMS and CWM. The average ratio and rms scatter of the small black circles that lie in one of the eight continuum bands of Table~2 are written in the panels for each of the nine A~stars.} 
\end{figure}

\begin{figure}
\plotone{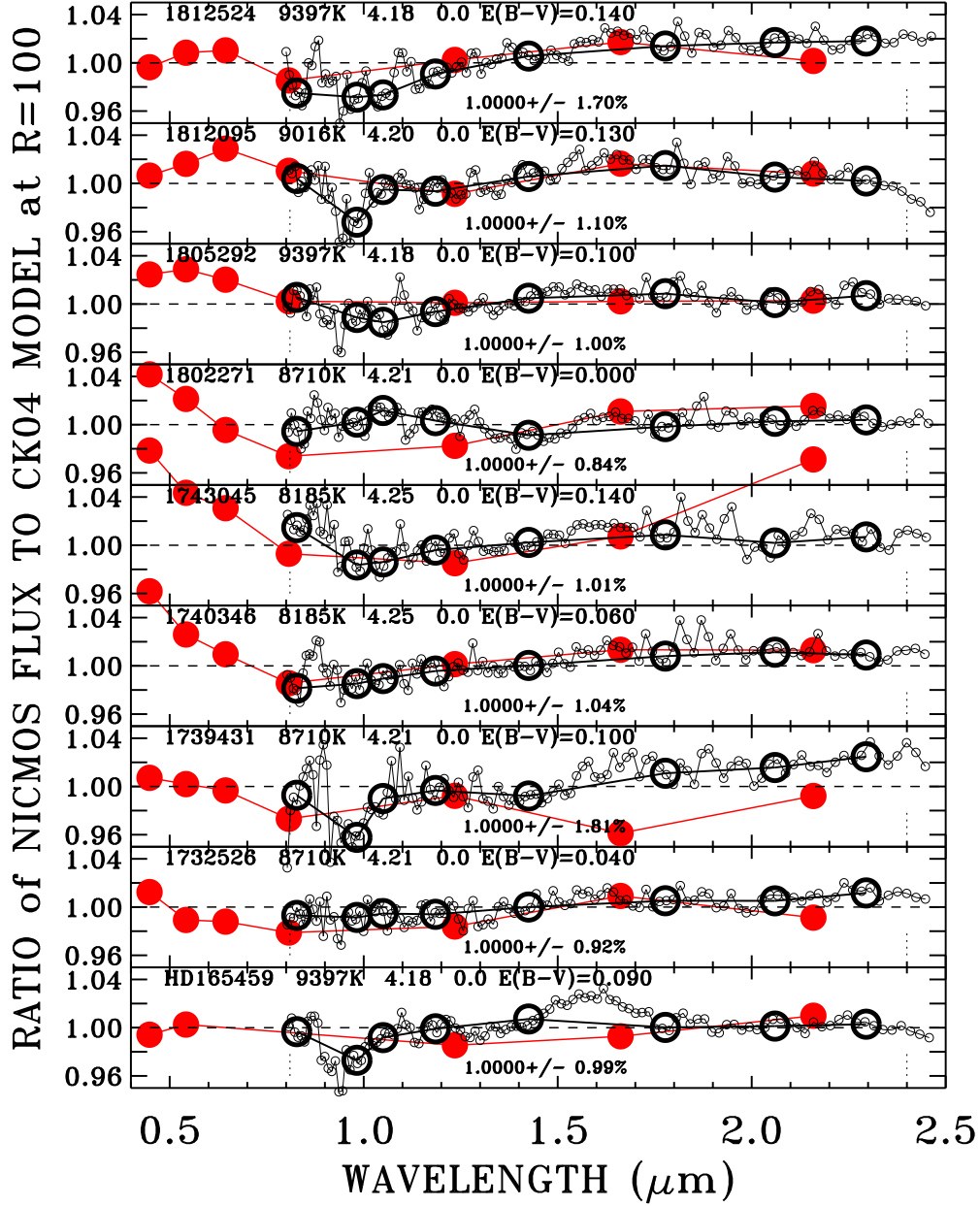}
\caption{
As in Figure 1, except that the denominators are interpolated from the CK04
model grid using the IDL routine \emph{ck04\_int.pro} provided by W.~Landsman. The stellar parameters for the CK04 models are written in each panel and are the same values as used for the CWW template models. The models are normalized to the NICMOS fluxes using the average ratio over the continuum regions of Table~2.} 
\end{figure}

\begin{figure}
\plotone{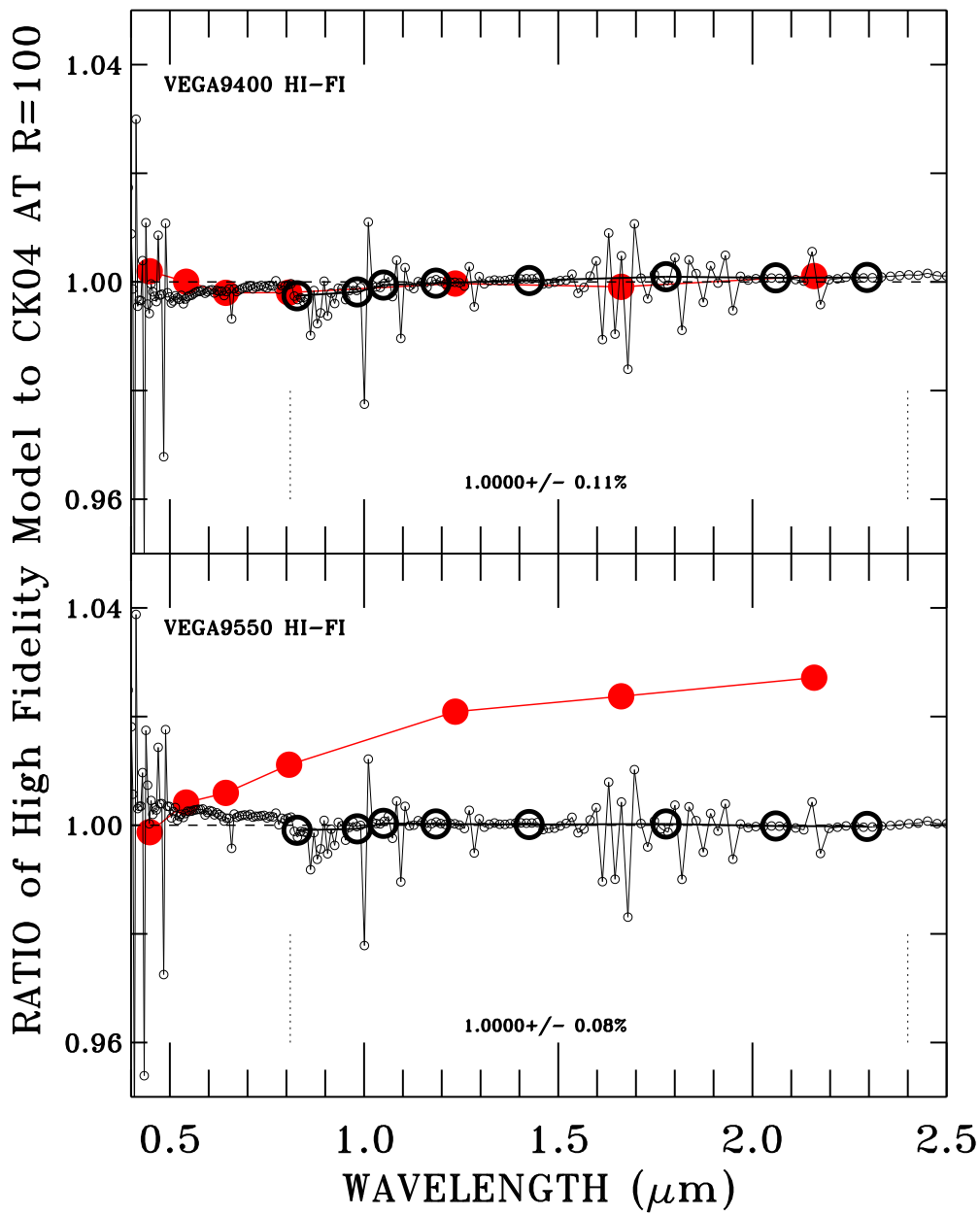}
\caption{
As in Figure 2, except that the numerators are high fidelity, special Kurucz
models for Vega with fine wavelength sampling. Both hi-fi models are normalized
to $3.46\times10^{-9}$~erg cm$^{-2}$ s$^{-1}$ at 5556~\AA. The zero points for the photometry (red filled circles) are established by the high fidelity 9400~K
Kurucz model. The excess scatter in the line regions excluded in Table~2 is
caused by the poorly sampled line profiles in the CK04 grid.}
\end{figure}

\begin{figure}  
\plotone{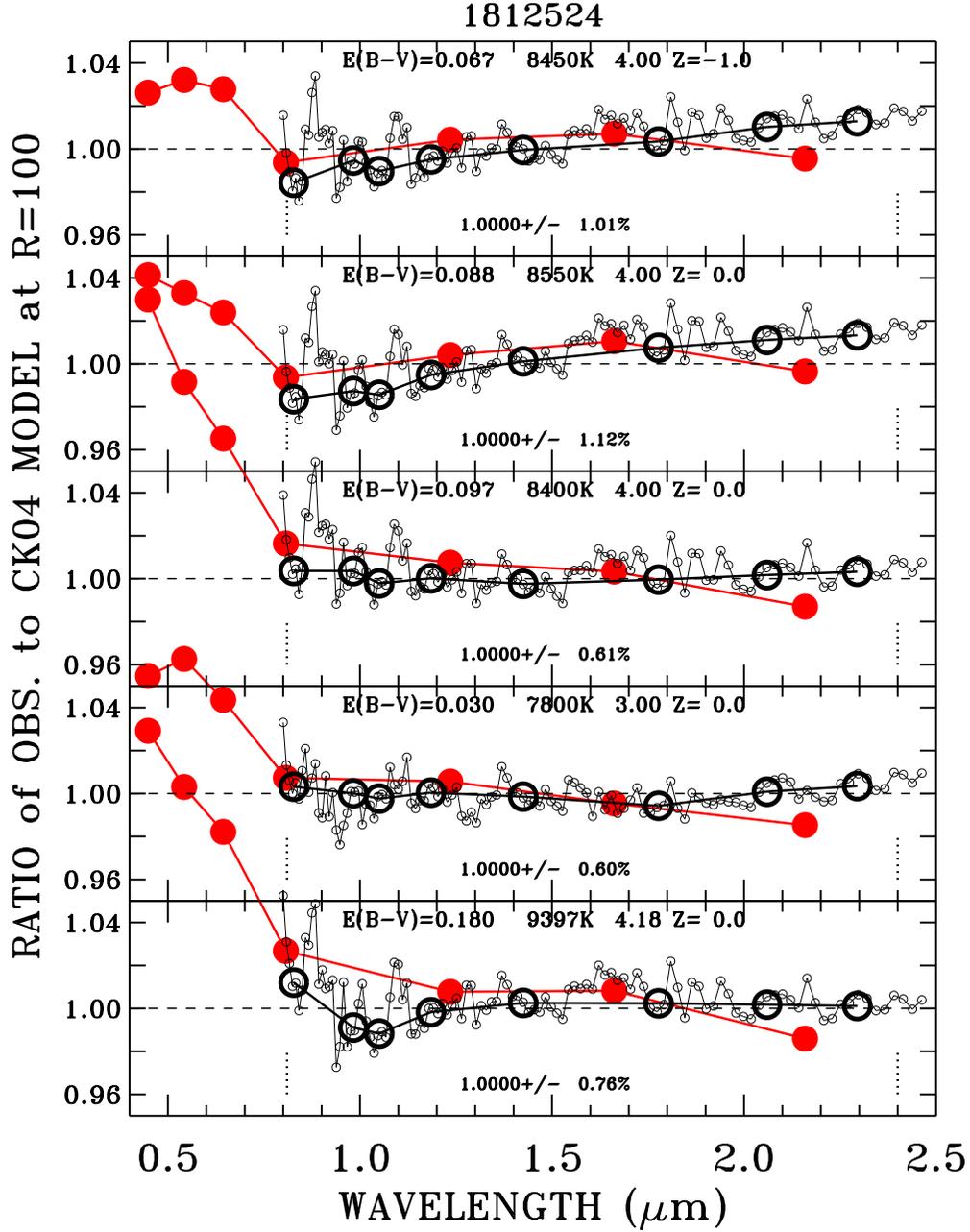}
\caption{
The bottom panel (a) repeats the top panel of Figure~2, except for the
interstellar reddening $E(B-V)$. The selective extinction $E(B-V)$ shown in
panels (a--c) minimizes the NICMOS rms scatter in the continuum bands for the
stellar parameters $T_\mathrm{eff}$, $\log g$, and $\log z$ that are also
written on each panel. Panel (b) shows the best fit to the NICMOS continuum.
Panel~(c) is the best fit for the constraint $\log g>4$, while panels (d--e)
also include the BVR photometry in the minimization of the rms scatter. In the
top panel, the denominator is the CK04 model which most closely matches the
measured SED with an overall rms minimum.} 
\end{figure}

\begin{figure}
\plotone{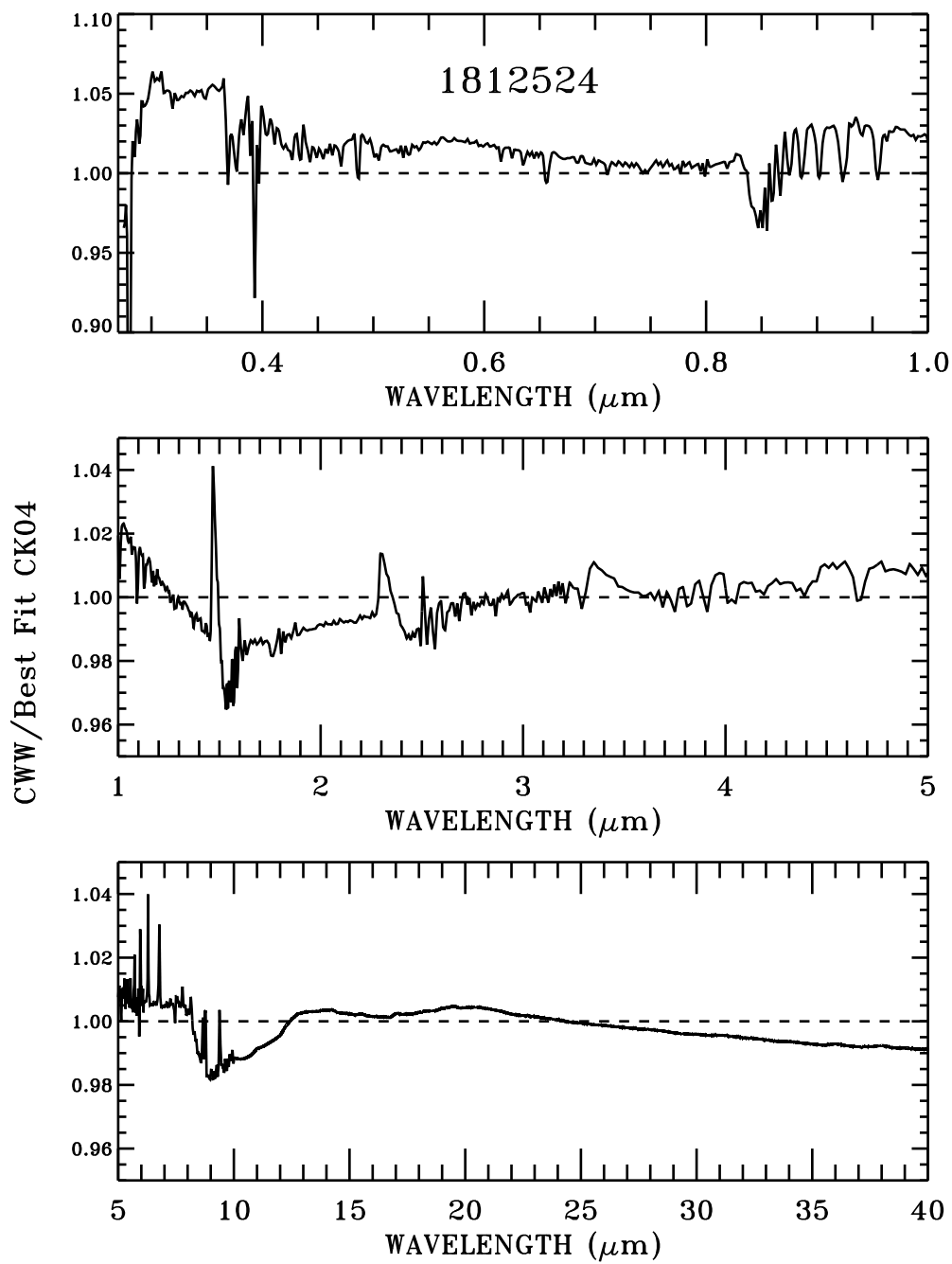}
\caption{
Ratio of CWW fluxes to a CK04 model normalized to the NICMOS data. The numerator is for the CWW stellar parameters $T_\mathrm{eff}=9397$, $\log g=4.18$, $\log z=0$, and $E(B-V)=0.14$, while the denominator is the best fit to the CWW photometry and NICMOS fluxes with $T_\mathrm{eff}=8450$, $\log g=4$, $\log z=-1$, and $E(B-V)=0.067$.}  
\end{figure}

\begin{figure}
\plotone{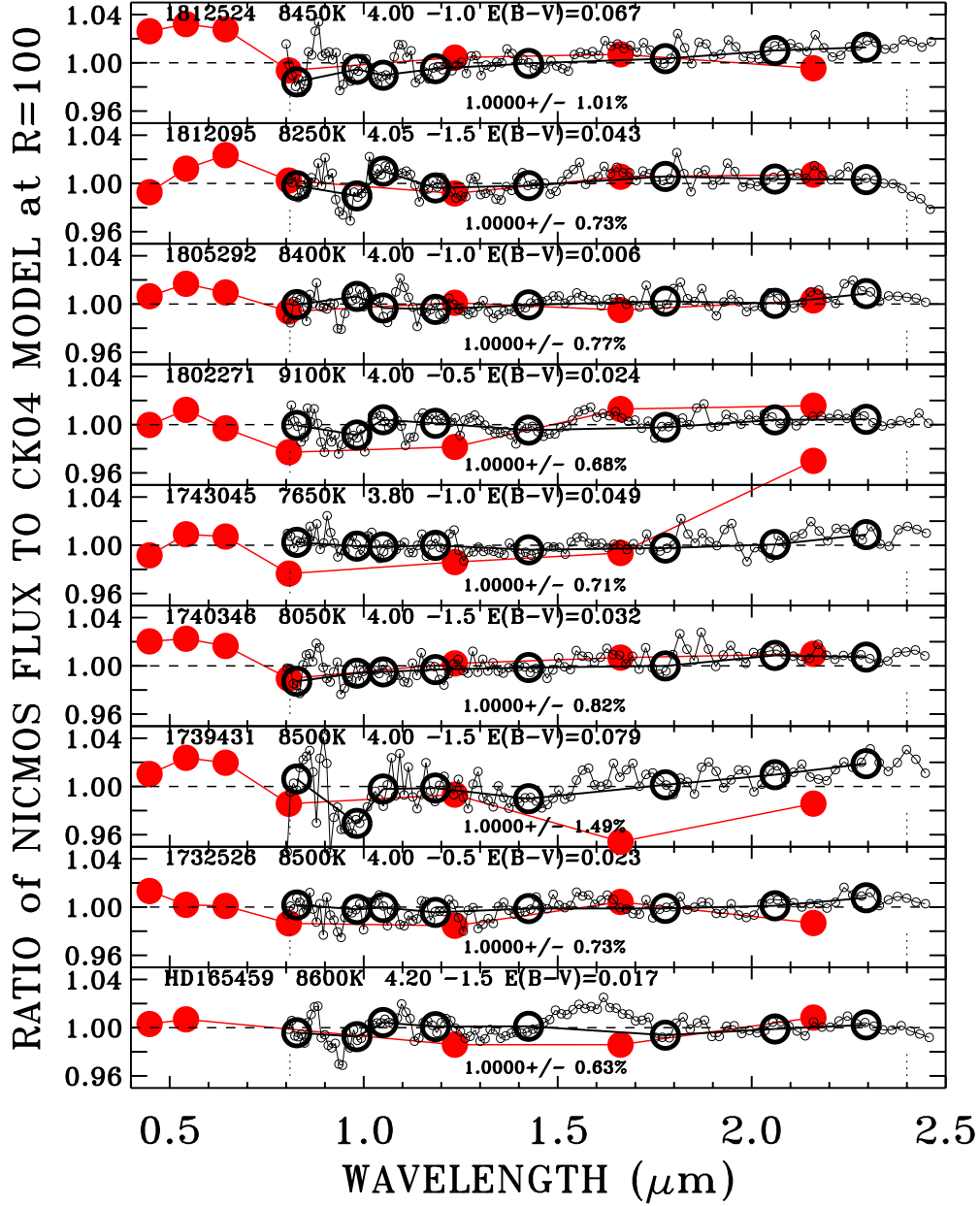}
\caption{
As in Figures 1 and 2, except that the denominators are the best fitting models
from the CK04 grid. The models are normalized to the NICMOS fluxes using the
average ratio in the continuum regions of Table~2. The values of the extinction
$E(B-V)$ minimize the residuals for the listed model atmosphere parameters. All
stars, except 1739431 and 1812524, have residual rms values within the expected
range.} 
\end{figure}

\end{document}